\documentclass{WileyMSP-template}
\usepackage{amsmath,amssymb,wasysym}
\usepackage{graphicx}
\usepackage{dcolumn}
\usepackage{bm}
\usepackage{braket}
\usepackage{subcaption}
\usepackage{verbatim}
\usepackage{float}
\usepackage{bbold}
\usepackage{listings}
\usepackage{color}
\usepackage{xcolor}
\usepackage{relsize}
\usepackage{amsthm}
\usepackage{siunitx}

\usepackage{enumerate}
\usepackage{accents}

\usepackage{soul,xcolor}
\usepackage[T1]{fontenc}
\usepackage[colorlinks=true ,urlcolor=blue,urlbordercolor={0 1 1}]{hyperref}

\newcommand{\beq}{\begin{eqnarray}}
\newcommand{\eeq}{\end{eqnarray}}

\usepackage{mathtools}

\newcommand{\bsp}{\begin{split}}
\newcommand{\esp}{\end{split}}

\newcommand{\bi}{\boldsymbol{i}}
\newcommand{\bj}{\boldsymbol{j}}

\graphicspath{{./figures/}}
\def\bea{\begin{eqnarray}}
\def\eea{\end{eqnarray}}

\makeatletter

\def\env@sqcases{%
  \let\@ifnextchar\new@ifnextchar
  \left\lbrack
  \def\arraystretch{1.2}%
  \array{@{}l@{\quad}l@{}}%
}
\makeatother
\definecolor{bleudefrance}{rgb}{0.19, 0.55, 0.91}

\begin{document}

\setstcolor{red}

\title{Fermionic spinon theory of the\\ hourglass spin excitation spectrum of the cuprates}

\maketitle{}

\author{Alexander Nikolaenko*}

\author{Pietro M. Bonetti}

\author{Subir Sachdev}


\begin{affiliations}
Alexander Nikolaenko, Pietro M. Bonetti, Subir Sachdev\\
Department of Physics, Harvard University, Cambridge MA 02138, USA\\
Email Address: onikolaienko@g.harvard.edu\\

Subir Sachdev\\
The Abdus Salam International Centre for Theoretical Physics, Strada Costiera 11, I-34151, Trieste, Italy.\\
Center for Computational Quantum Physics, Flatiron Institute, 162 5th Avenue, New York, NY 10010, USA
\end{affiliations}

\begin{abstract}
We present a theory for the spin fluctuation spectrum of the hole-doped cuprates in a ground state with period 4 unidirectional charge density wave (`stripe') order. Motivated by recent experimental evidence for a fractionalized Fermi liquid (FL*) description of the intermediate temperature pseudogap metal, we employ a theory of fermionic spinons which are confined with the onset of stripe order at low temperatures. The theory produces the   `hourglass' spectrum near stripe-ordering wavevector observed by neutron scattering. Additional scattering from spinon continua and bound states appears at higher energies and elsewhere in the Brillouin zone, and could be observed by neutron or X-ray scattering.
\end{abstract}


\section{Introduction}
\label{sec:intro}

There have been a number of earlier studies suggesting the relevance of fractionalized spinon excitations (which carry spin 1/2 and electrical charge 0) to square lattice antiferromagnets \cite{Hayden10,Ronnow15,Becca18,Ma2018,Yasir19,Ronnow25}. Recent work \cite{Mascot2022,PNAS_pseudo1,Christos2024,BCS24,Sayantan25}, reviewed in Refs.~\cite{Boulder25,SSLec25}, has argued that such spinons are the key to understanding a number of observations in the underdoped cuprates. These arguments rely on the fractionalized Fermi liquid (FL*) theory \cite{FLS1,Si2004,FLS2,APAV04,GroverSenthil10,Bonderson16,Tsvelik16,Vojta20,ElseSenthil,Tsvelik24} of the pseudogap phase: the FL* state of the doped square lattice antiferromagnet \cite{Kaul07,ACL08,Qi10,Punk15,Zhang2020,Mascot2022} has 4 Fermi pockets of spin-1/2 charge $e$ quasiparticles, each of fractional area $p/8$, so that the total density of mobile quasiparticles is $p$. Such a state does not obey the Luttinger Fermi surface area \cite{Luttinger60}, but does satisfy the Oshikawa anomaly \cite{MO00} by the presence of additional spinon excitations \cite{FLS1,FLS2,APAV04,Bonderson16,ElseSenthil}. 

Compelling evidence for such a FL* state comes from recent angle-dependent magnetoresistance (ADMR) measurements \cite{fang_admr_2022,chan_yamaji_2024}. The ADMR displays significant evidence for the presence of hole pockets which can tunnel coherently between square lattice layers, and the Yamaji effect is consistent with the FL* pocket area of $p/8$ \cite{Zhao_Yamaji_25}. Moreover, the Yamaji effect requires coherent inter-layer electronic transport, which is present for FL*, and unlikely for other pseudogap candidates \cite{Zhao_Yamaji_25}. Other recent arguments for the FL* description of the pseudogap are summarized in Section~\ref{sec:conclusions}.

In this paper, we turn to the role of spinons in the spin-fluctuation spectrum, where they have a direct role. Our focus is on the well-studied `hourglass' spectrum of the stripe state observed by neutron scattering in the cuprates near hole doping $p=1/8$ \cite{Hayden2004,Vignolle2007,Chan2016}. The name `hourglass' comes from the shape of the spectrum, which has a distinctive narrowing in the middle and widening at higher and lower energies.

We use the $\pi$-flux spin liquid theory for the underlying square lattice antiferromagnet \cite{AM88,Affleck-SU2,Fradkin88}. This has fermionic spinon excitations which carry spin $S=1/2$ and a dispersion with 2 massless Dirac nodes in the square lattice Brilloin zone. The spinons are coupled to an emergent SU(2) gauge field (the gauge SU(2) is distinct from, and commutes with, the spin rotation SU(2)). Fluctuations of the SU(2) gauge field in the insulator are expected to lead to confinement of the spinons into either the N\'eel state or the valence bond solid (VBS) state, but there is evidence for a deconfined critical state of the spinons between the N\'eel and VBS states \cite{NRSS89prl,Wang17,Fuzzy24}. The free spinon dynamic structure factor is reviewed in Section~\ref{sec:freespinons}, and this is in good correspondence with numerical studies of the N\'eel-VBS transition \cite{Becca18}. (Ref.~\cite{Becca18} employs a $\mathbb{Z}_2$ spin liquid for their numerics, but this spin liquid is closely connected to the SU(2) spin liquid \cite{Shackleton21}, and has a similar spinon spectrum.) The free spinon theory exhibits significant low energy spectral weight near momenta $(\pi, \pi)$, $(\pi, 0)$, and $(0,0)$; Ref.~\cite{BCS24} argued that SU(2) gauge fluctuations strongly suppress the spectral weight near momenta $(0,0)$ and $(\pi, 0)$.

Resonant inelastic X-ray scattering (RIXS) experiments have investigated the high energy spin fluctuation spectrum in the cuprates \cite{Keimer11,Hayden19,Keimer25}. Ref.~\cite{SSZaanen} has noted that the broad continuum scattering between momenta $(0,0)$ and $(\pi,0)$ bear a strong resemblance to the continuum spectral weight in the frustrated square lattice antiferromagnet near the N\'eel-VBS transition \cite{Becca18}. The energy scales match well, and the main difference is that the insulating antiferromagnet has gapless spectral weight at $(\pi, 0)$, while the doped cuprate has little spectral weight below $\sim 200$ meV. This difference could be a consequence of the doped carriers, whose influence will be studied here.

The present paper will study a different route to confinement of the spinons at low energies, appropriate to the doped cuprates \cite{PNAS_pseudo1}. We will assume a charge density wave (CDW) with static long-range order which is produced by the condensation of a Higgs field $B$ \cite{BCS24}. This Higgs field carries a fundamental SU(2) gauge charge, and so its condensation gaps out all components of the SU(2) gauge field. The resulting $B$-condensed phase is `topologically trivial', but nevertheless, signatures of the deconfined spinons can emerge at higher energies. And at low energies, as we shall demonstrate, such a spinon theory is able to reproduce the `hourglass' spectrum of the CDW (`stripe') state. Our approach differs from other theories of the hourglass \cite{Vojta2004,Vojta2006,Vojta2012,Vojta2013} by using spin $S=1/2$ excitations in a background of charge order, whereas the earlier works considered spin $S=1$ paramagnon excitations.

We study the spin susceptibility of the underdoped cuprates using the Ancilla Layer Model (ALM) studied extensively in previous works \cite{Zhang2020,Zhang2020_2,Nikolaenko2023,Christos2024,BCS24,Boulder25}. The ALM offers a unified description of both the FL* pseudogap metal with small Fermi surface, and a conventional Fermi liquid with large Fermi surface. It also hosts fractionalized spinon excitations that ensure consistency with the Oshikawa anomaly. In the following sections, we examine how these excitations affect the spin structure factor and make predictions for future neutron scattering experiments. 

In Section \ref{sec:theory} we introduce the ALM, and derive the expressions for susceptibility in the presence of CDW order. In Section \ref{sec:Discussion} we present our calculation of the spin structure factor, discuss the role of spinons and compare our calculations to experiments. Finally, we conclude in Section \ref{sec:conclusions} by recalling experimental and theoretical evidence for FL* in the hole-doped cuprates.

\section{Theory}
\label{sec:theory}
We start this section by formulating the ALM. The model has three layers: the first level consists of electrons described by tight-binding model,  while the second and the third layers are filled with spins with Heisenberg interaction between them.
The Hamiltonian of the ALM is \cite{Zhang2020,Mascot2022,Christos2024}:
\begin{equation}
   H= \sum_{\bi,\bj} \left\{-t_{\bi\bj} c^\dagger_{\bi,\alpha} c_{\bj,\alpha}+J_{1,\bi\bj}\, {\bm S}_{1,\bi}\cdot {\bm S}_{1,\bj}+J_{2,\bi\bj}\, {\bm S}_{2,\bi}\cdot {\bm S}_{2,\bj}\right\}\,+ \sum_{\bi}\left\{ J_K {\bm S}_{c,\bi}\cdot {\bm S}_{1,\bi}+J_\perp {\bm S}_{1,\bi}\cdot {\bm S}_{2,\bi}\right\}\,,
\label{eq:initial_Hamiltonian} 
\end{equation}
where $\bi,\bj \in \Lambda$, with $\Lambda=\{(x,y)\,|\,x,y\in\mathbb{Z}\}$ being a two-dimensional square lattice. 
For the first layer we employ the tight-binding parameters $t=0.22$ (nearest neighbor hopping), $t'=-0.034$ (second neighbor hopping), $t''=0.036$ (third neighbor hopping), $t'''=-0.007$ (fourth neighbor hopping), and a chemical potential $\mu_c=-0.24$ (all in units of eV). As in Ref.~\cite{Mascot2022}, these values have been extracted by fitting photoemission data of Ref.~\cite{He2011} in the overdoped regime, where a large Fermi surface is clearly visible. Indeed, in the overdoped phase, the second and third layers of the ALM effectively decouple from the rest of the system by forming rung singlets. Thus, the physics is described by the first layer only, which is a Fermi liquid with a large Fermi surface, whose shape is determined by the hoppings $t_{\bi\bj}$, as in conventional tight-binding models.

We use the Schwinger-fermion representation for the second ($a=1$) and third ($a=2$) layer spins:
$ {\bm S}_{a,\bi} =(1/2)  f_{a,\bi, \alpha}^\dagger {\bm \sigma}_{\alpha\beta} f_{a,\bi, \beta}$ with the imposed constraint $\sum_{\alpha} f_{a,\bi,\alpha}^\dagger f_{a,\bi,\alpha} = 1$, satisfied at every lattice site $i$. 
Throughout the paper, we adopt a mean-field approach and neglect all gauge field fluctuations emerging from the redundancy of the parton construction. Within the mean-field theory, the interacting terms in the Hamiltonian simplify to:
\begin{equation}\label{eq:middle bottom layers decoupled}
H_{f_1,f_1}+H_{c,f_1}+H_{f_1,f_2}= t_{1,\bi\bj} f^\dagger_{1,\bi,\alpha}f_{1,\bj,\alpha} +\Phi (f_{1,\bi, \alpha}^\dagger c_{\bi,\alpha}+c_{\bi,\alpha}^\dagger f_{1,\bi, \alpha}) +\left[i B_{1,\bi} f_{2,\bi, \alpha}^\dagger f_{1,\bi,\alpha} + i B_{2,\bi} f_{2,\bi,\alpha}\varepsilon_{\alpha\beta} f_{1,\bi,\beta}+\text{h.c.} \right] \,, 
\end{equation}

where $\Phi\propto J_K \langle f_{1,\bi, \alpha}^\dagger c_{\bi,\alpha} \rangle \in\mathbb{R}$ and $B_{1,\bi} \propto J_\perp \langle f_{1,\bi, \alpha}^\dagger f_{2,\bi,\alpha} \rangle$, $B_{2,\bi} \propto J_\perp \langle f_{1,\bi, \alpha}^\dagger\varepsilon_{\alpha\beta} f^\dagger_{2,\bi,\beta} \rangle$, with $\varepsilon_{\alpha\beta}$ the unit antisymmetric tensor. The second layer is characterized by the hopping parameters $t_1=0.1, t_1'=-0.03, t_1''=-0.01$, a chemical potential $\mu_f=0.009$, and a hybridization with the first layer $\Phi=0.09$.
The finite hybridization between the first and second layers leads to the emergence of four Fermi pockets in the Brillouin zone with area $p/8$. Moreover, the intensity at back side of the pockets is subdued, which is consistent with Fermi arcs features observed in photoemission experiments. The parameters were chosen to reproduce the photoemission data for Bi2201 at a doping level of $p=0.206$~\cite{Mascot2022,He2011}. The $B$ field would be discussed later, to incorporate the effects of CDW order. For now we assume $B_{a,\bi}=0$.

A nonzero value of $\Phi$ reconstructs the top layer large Fermi surface, of (hole) volume $1+p$, into four hole pockets of total volume $p$. Because no translation symmetry breaking is occurring at this point, this is at odds with the conventional formulation of Luttinger's theorem. To account for its violation, we assume the presence of a quantum spin liquid in the bottom layer, hosting fractionalized spinon excitations that account for the missing Luttinger volume~\cite{MO00,FLS1,FLS2}. In particular, we choose a $\pi$-flux spin liquid Ansatz of the form:
\begin{equation}\label{eq:H_pi_flux}
    H_{f_2,f_2}=i t^{f_2}f^\dagger_{2,\bi,\alpha}e_{\bi,\bj}f_{2,\bj,\alpha}\,,
\end{equation}
with $e_{\bi,\bj}=-e_{\bj,\bi}$ and in the chosen gauge $e_{\bi,\bi+\hat{\boldsymbol{x}}}=1$, $e_{\bi,\bi+\hat{\boldsymbol{y}}}=(-1)^x$. We also set $t^{f_2}=0.14 \, eV$ similar to Ref. \cite{Christos2024}. The chosen gauge is not manifestly translation invariant in the $x$-direction. In fact, translations along the $x$-axis are implemented projectively $f_{2,\bi,\alpha}\to (-1)^y f_{2,\bi,\alpha}$, implying that all gauge invariant quantities computed with Hamiltonian~\eqref{eq:H_pi_flux} will be translation invariant. The Hamiltonian $H_{f_2,f_2}$ can be diagonalized in the momentum space. We consider a supercell with two sites ($\delta_{\bi}=0,1$) in the $x$-direction. After Fourier transform $f_{2,\delta_i}(\bm{r})=e^{i \bm{k} \bm{r}}f_{2,\delta_i}(\bm{k})$ the Hamiltonian in the basis 
 $\{f^\dag_{2,\delta_i}(\bm{k})\}$ is $2 \times 2$ matrix with with spectrum $E_{\pm}(k)=\pm 2 t^{f_2} \sqrt{\sin(k_x )^2+\sin(k_y)^2 }$ and $\bm{k}$ is defined in the \textit{reduced} Brillouin zone $\mathrm{BZ}'=\{ (k_x,k_y)\,|\, k_x\in(-\pi/2,\pi/2),\, k_y\in(-\pi,\pi) \}$. The spectrum features two Dirac cones (per spin projection $\alpha$) at $(k_x,k_y)=(0,0)$ and $(k_x,k_y)=(0,\pi)$.

\subsection{Bare spin susceptibilities for the $\pi$-flux spin liquid}
\label{sec:freespinons}

The bare $\pi$-flux spin-$z$ susceptibility in real space is given by the product of two Green functions:
\begin{equation}
\chi^0_{\bi,\bj}(i\Omega_n)=\sum_{m=-\infty}^{+\infty} G_{\bi,\bj}(i \omega_m)G_{\bj,\bi}(i \omega_m+i \Omega_n) \,,
\label{eq:bare_susc}
\end{equation}
with $\omega_m$ and $\Omega_n$ fermionic and bosonic Matsubara frequencies, respectively, and $G_{\bi,\bj}(i\omega_n)=[i\omega_n-i t^{f_2}e_{\bi,\bj}]^{-1}$.
As previously discussed, being $\chi^0_{\bi,\bj}(i\Omega_n)$ a gauge invariant object, we expect it to be translation invariant, that is, $\chi_{\bi,\bj}(i\Omega_n)=\chi_{\bi-\bj}(i\Omega_n)$. We define $\pi$-flux susceptibility in the momentum basis as 
\begin{equation}
  \chi^0(\omega,\bm{q})=\sum_{\bi,\bj}\chi^0_{\bi,\bj}(i \Omega_n \rightarrow\omega+i 0^+)e^{-i \bm{q}\cdot (\bi - \bj)} .
  \label{eq:bare_susc_exp}
\end{equation}

After analytically continuing to real frequencies and going to momentum basis, we obtain:
\begin{equation}
   \chi^0(\omega,\bm{q})=\sum_{\ell,\ell'=\pm} \sum_{\bm{k}} \frac{n_F(E_\ell(\bm{k}))-n_F(E_{\ell'}(\bm{k}+\bm{q}))}{\omega+i 0^++E_\ell(\bm{k})-E_{\ell'}(\bm{k}+\bm{q})}| \langle E_{\ell'}(\bm{k}+\bm{q})|E_\ell(\bm{k})\rangle |^2, 
\end{equation}
with $E_\alpha(\bm{k})$ and $\ket{E_\alpha(\bm{k)}}$ denoting an eigenvalues and eigenvectors of the Hamiltonian in Eq.~\eqref{eq:H_pi_flux}, $\sum_{\bm{k}}$ is a shorthand for $\int_{\bm{k}\in\mathrm{BZ}'}\frac{d^2\bm{k}}{(2\pi)^2}$, and $n_F(x)$ is the Fermi function.

\subsection{Bare susceptibilities in the presence of charge density wave order}

We now consider the susceptibilities when $B_{a,\bi}\neq 0$. As found in Ref.~\cite{PNAS_pseudo1}, the condensation of the field $B_{a,\bi}$ has both effects of fully \textit{Higgsing} the SU(2) gauge group associated with the Schwinger fermion decomposition in the bottom layer, and breaking some physical global symmetries, such as translations, time reversal, or U(1) charge conservation, depending on the spatial pattern of $B_{a,\bi}$. In the following, we choose an \textit{Ansatz} for $B_{a,\bi}$ that only breaks translation symmetry along the $x$-axis:
\begin{equation}
    B_{1,\bi}=2b\,i^{x+y}[\cos \theta+(-1)^x \sin \theta]\,\cos\left(\frac{K_xx+\phi}{2}\right)\,, \quad B_{2,\bi}=0\,,
    \label{eq:B_field}
\end{equation}
with  $b \in \mathbb{R}$ denoting the amplitude of the wave and $\phi$ a phase shift.

According to~\cite{PNAS_pseudo1}, the CDW order parameter is given by:
\begin{equation}
    \rho_{\bi}=B_{\bi}^\dagger B_{\bi}= 2b^2[1+(-1)^x \sin 2\theta ][\cos\left(K_x x+\phi\right)+1]\,.
\end{equation}
One can check that the above \textit{Ansatz} gives vanishing current and superconducting order parameters, $\Re[B^\dagger_{\bi} e_{\bi,\bj} B_{\bj}]=0$, $B_{a,\bi}\varepsilon_{ab}e_{\bi,\bj} B_{b,\bj}=0$. In the following, we restrict ourselves to a period-4 bond-centered CDW, given by $K_x=\frac{\pi}{2}$, $\phi=\frac{\pi}{4}$. Such CDW order is well established in some cuprate materials, for example optimally doped LBCO \cite{Tranquada1995,Fujita2004}.

We consider two different effects of the $B_{a,\bi}$ condensate on the fermionic degrees of freedom: one is given the last terms in Eq.~\eqref{eq:middle bottom layers decoupled}, which couple the bottom layer spinons to the top two layers, making them indistinguishable from physical electrons; a second effect is that it could modify the $\pi$-flux \textit{Ansatz} into
\begin{equation}
    H_{f_2,f_2}=i t^{f_2}(1+\lambda Q_{\bi,\bj})f^\dagger_{2,\bi,\alpha}e_{\bi,\bj}f_{2,\bj,\alpha},
    \label{eq:H_pi_flux_CDW}
\end{equation}
with 
\begin{equation}
\begin{split}
    & Q_{\bi,\bi+\hat{\boldsymbol{x}}}=\Im [B^\dagger_{\bi} e_{\bi,\bi+\hat{\boldsymbol{x}}} B_{\bi+\hat{\boldsymbol{x}}}]=2b^2\cos 2 \theta \,[\cos (K_x(x+1/2)+\phi)+\cos(K_x/2) ]\,,\\
       &  Q_{\bi,\bi+\hat{\boldsymbol{y}}}=\Im [B^\dagger_{\bi} e_{\bi,\bi+\hat{\boldsymbol{y}}} B_{\bi+\hat{\boldsymbol{y}}}]=2 b^2[(-1)^x+\sin 2 \theta][\cos \left(K_x x+\phi \right)+1]\,.
\end{split}
\end{equation}
Note that this last term cannot be derived with the sole use of mean-field theory, but it can be thought of as emerging upon integrating out of some high-energy degrees of freedom. Similarly to the previous case, we introduce a unit cell consisting of four sites $\delta_i=0,1,2,3$ with 
basis $\{f^\dag_{2,\delta_i}(\bm{k})\}$ and the following  Fourier transform between real and Fourier space:  $f_{2,\delta_i}(\bm{r})=e^{i \bm{k} \bm{r}}f_{2,\delta_i}(\bm{k})$ The Hamiltonian in Eq.~\eqref{eq:H_pi_flux_CDW} becomes a $4\times 4$ matrix. On site constraint $ \langle f^\dag_{2,i} f_{2,i}\rangle=1$ is also satisfied since the Hamiltonian remains particle-hole symmetric. Eq.~\eqref{eq:bare_susc} is readily generalized in the new basis:

\begin{equation}
   \chi_{i,j}^0( i q_n) =\sum_{\omega_n,k,q}G_{\delta_i \delta_j}(i \omega_n+i q_n, \bm{k}+\bm{q})G_{\delta_j,\delta_i}(i \omega_n, \bm{k})e^{i \bm{q}(i-j)}= \chi^0_{\delta_i \delta_j}(i q_n, \bm{q})e^{i \bm{q} (i-j)}.
   \label{eq:suscept_CDW}
\end{equation}

After analytically continuing to real frequencies, we obtain:
\begin{equation}
    \chi^0_{\delta_i,\delta_j}(\omega,\bm{q})=\sum_{\alpha,\beta=1,..,4}\sum_{\bm{k}} \frac{n_F(E_\alpha(\bm{k}))-n_F(E_\beta(\bm{k}+\bm{q}))}{\omega+i \delta+E_\alpha(\bm{k})-E_\beta(\bm{k}+\bm{q})}  F^\alpha_{\delta_i}(\bm{k}) F^{\alpha}_{\delta_j}(\bm{k})^*  F^\beta_{\delta_j}(\bm{k}+\bm{q})F^\beta_{\delta_i}(\bm{k}+\bm{q})^* ,
    \label{eq:susc:CDW}
\end{equation}
with the form factors $F^\alpha_{f_2,\delta_i}(\bm{k})=\bra{f_2,\delta_i}\ket{E_\alpha(\bm{k})} $ and $\ket{E_\alpha(\bm{k})} $ being eigenvectors of the Hamiltonian in Eq.~\eqref{eq:H_pi_flux_CDW}. 

Now, the interacting part of the Hamiltonian is:
\begin{equation}
    H_{int}=\sum_{\bi\bj}J_{2,\bi\bj}(1+g Q^b_{\bi\bj})f^\dag_{\bi\alpha} f_{\bj \alpha} f^\dag_{\bj \beta} f_{\bi \beta}=\sum_{\bi\bj}\tilde{J}_{\bi\bj}f^\dag_{\bi\alpha} f_{\bj \alpha} f^\dag_{\bj \beta} f_{\bi \beta},
    \label{eqn:Ham_int}
\end{equation}
where $J_{2,\bi\bj}$ are the Heisenberg couplings in the bottom layer. We choose $J_{2,\bi\bj}$ so that its Fourier transform $J_2(\bm{q})$ is given by 
\begin{equation}
    \begin{split}
        J_2(\bm{q}) = &2J\left(\cos q_x+\cos q_y\right) + 4J'\cos q_x \cos q_y + 2J^{\prime\prime}\left(\cos 2q_x+\cos 2q_y\right) \\&+ 4J^{\prime\prime\prime}\left(\cos 2q_x\,\cos q_y+\cos q_x\,\cos 2q_y\right)\,.
    \end{split}
\end{equation}
We will include RPA corrections from this interaction, following the application of such methods to the triangular lattice antiferromagnet \cite{Batista18,LiZhang20,Knolle25}.
After using the Fourier representation 
\begin{equation}
    J_{ij}=\sum_q J_{\delta_i \delta_j}(\bm{q}) e^{i \bm{q} (i -j)}\,,    
\end{equation}
and summing up the series, the RPA result in the spin-spin channel in matrix notation reads:
\begin{equation}
    \chi^{RPA}(\omega,\bm{q})=\chi^0(\omega,\bm{q})(1-\tilde{J}(\bm{q})\chi^0(\omega,\bm{q}))^{-1}.
    \label{eq:susceptibility_RPA}
\end{equation}

Although $\chi^{RPA}(\omega,\bm{q})$ is $4 \times 4$ matrix, Eq.~\eqref{eq:bare_susc_exp} provides an efficient route to extract an experimentally relevant expression. Noting that $\chi^{RPA}_{\delta_i,\delta_j}(\omega,\bm{q}+n K_x)e^{i n K_x (\delta_i-\delta_j)}=\chi^{RPA}_{\delta_i,\delta_j}(\omega,\bm{q})$ and accounting for charge density wave (CDW) order along both spatial directions, the experimentally observable susceptibility can be written as: $\chi^{exp}(\omega, \bm{q})=\sum_{\delta_i,\delta_j} \left(\chi^{RPA}_{\delta_i,\delta_j}(\omega,q_x,q_y)+\chi^{RPA}_{\delta_i,\delta_j}(\omega,q_y,q_x)\right)$.
For the expression of the Hamiltonian, and $\tilde{J}$ in the matrix notation see Appendix \ref{app:CDW}.

The full Hamiltonian, including couplings to electrons, could be written as:
\begin{equation}
H= \psi_{\bm{k}}^*
\left(
    \begin{array}{ccc}
H_{c,c}(\bm{k})&\Phi^\dag&0\\
\Phi & H_{f_1, f_1}(\bm{k})&-i g_eB^\dag\\
0 &i g_e B&H_{f_2,f_2}(\bm{k})\\
    \end{array}
    \right)    \psi_{\bm{k}},
\label{eq:hamiltonian_electrons}
\end{equation}
where $\psi_{\bm{k}}=(c_{\bm{k},\delta_1,\uparrow},\dots, c_{\bm{k},\delta_4,\uparrow},f_{1,\bm{k},\delta_1,\uparrow},\dots, f_{1,\bm{k},\delta_4,\uparrow},f_{2,\bm{k},\delta_1,\uparrow},\dots, f_{2,\bm{k},\delta_4,\uparrow})$ consists of three layers in the extended CDW basis and $g_e$ determines the strength of hybridization between the third and second layers.

The bare susceptibility is determined analogous to Eq.~\eqref{eq:suscept_CDW}:
\begin{equation}
      \chi^{f_2, f_2}_{\delta_i,\delta_j}(i q_n,\bm{q})^0=\sum_{\bm{k}, i \omega_n}G_{\delta_i,\delta_j}^{f_2 f_2} (i q_n+i\omega_n,\bm{k}+\bm{q})G^{f_2 f_2}_{\delta_j, \delta_i}(i\omega_n,\bm{k})
\end{equation}
And after performing Matsubara summation and analytically continuing to real frequencies:

\begin{equation}
    \chi^{ f_2 f_2}_{\delta_i,\delta_j}(\omega,\bm{q})^0=\sum_{\alpha,\beta=1,..,12}\sum_{\bm{k}} \frac{n_F(E_\alpha(\bm{k}))-n_F(E_\beta(\bm{k}+\bm{q}))}{\omega+i \delta+E_\alpha(\bm{k})-E_\beta(\bm{k}+\bm{q})}  F^\alpha_{f_2,\delta_i}(\bm{k}) F^{\alpha}_{f_2,\delta_j}(\bm{k})^*  F^\beta_{f_2,\delta_j}(\bm{k}+\bm{q})F^\beta_{f_2,\delta_i}(\bm{k}+\bm{q})^* ,
    \label{susc:electrons}
\end{equation}
with the form factors $F^\alpha_{f_2,\delta_i}(\bm{k})=\bra{f_2,\delta_i}\ket{E_\alpha(\bm{k})} $ and $\ket{E_\alpha(\bm{k})} $ being eigenvectors of the Hamiltonian in Eq.~\eqref{eq:hamiltonian_electrons}. Eqs.~(\ref{eq:susceptibility_RPA},\ref{susc:electrons}) provide the most general expressions for the susceptebility and we use them in the next section to analyze the spin structure factor near optimal doping.

\section{Discussion}
\label{sec:Discussion}

\begin{figure}[h]
\begin{minipage}[h]{1\linewidth}
  \center{\includegraphics[width=1\linewidth]{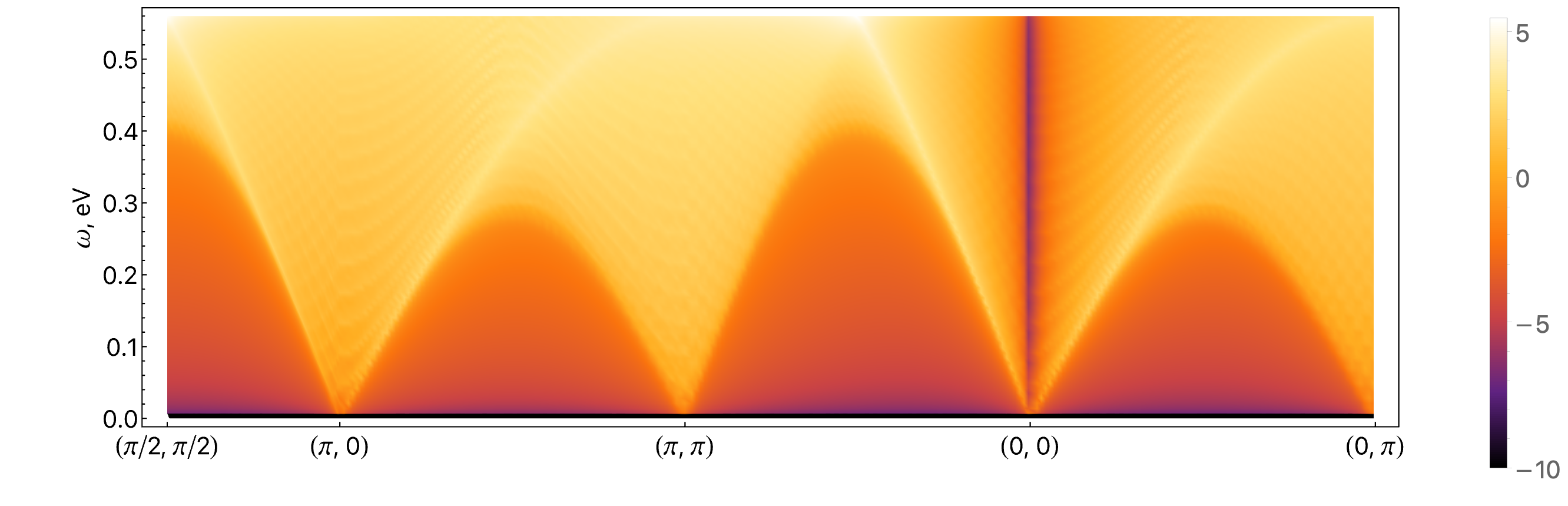}}
  \end{minipage} 
\caption{Bare $\pi$-flux spin structure factor $-\Im \chi^0(\omega,q)$ on a logarithmic scale along high-symmetry directions in the Brillouin zone. The regions of strong intensity correspond to a spinon continuum, having a Dirac-like shape near $(\pi,\pi)$. }
\label{fig:susceptibility_barepiflux}
\end{figure}

In this section, we analyze the spin-structure factor obtained from our model and compare it with experimental observations in various cuprate compounds. 
We assume that the dominant contribution to the spin-structure factor comes from the third layer which is a $\pi$-flux spin liquid described by the Hamiltonian in Eq.~\eqref{eq:H_pi_flux}. The system is treated within the mean-field approximation, neglecting the SU(2) fluctuations of the gauge field.
Fig.~\ref{fig:susceptibility_barepiflux} shows the imaginary part of the spin susceptibility for the third layer. A similar calculation was previously done in Ref.~\cite{Ma2018}. As shown, there is a gapless spinon continuum centered at $(\pi,\pi)$ in momentum space, exhibiting a Dirac-like dispersion.


\begin{figure}[h!]
\begin{minipage}[h]{1\linewidth}
  \center{\includegraphics[width=1\linewidth]{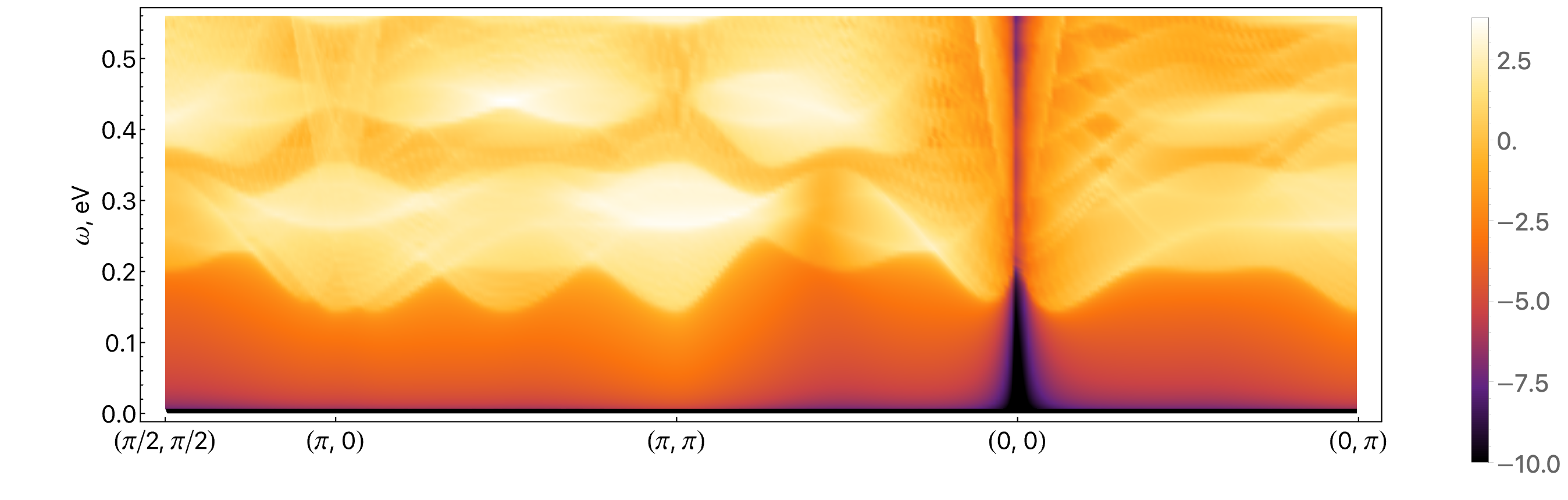}}
  \end{minipage} 
\caption{Bare $\pi$-flux spin structure factor $-\Im \chi^0(\omega,q)$  on a logarithmic scale in the presence of CDW order, averaged over the $x$ and $y$-directions. As a result of CDW order, the gap opens throughout the Brillouin zone, moving the spinon continuum to higher energies. }
\label{fig:susceptibility_barepiflux_CDW}
\end{figure}

As discussed in the previous section, the condensation of the $B$-field Higgs the SU(2) gauge field and leads to CDW order. We assume that the CDW order is bond-centered and has a period of four unit cells. The CDW potential opens a gap in the $\pi$-flux Dirac-like spectrum and moves a spinon continuum to higher energies. This gap directly manifests in the spin-structure factor, as illustrated in Fig.~\ref{fig:susceptibility_barepiflux_CDW}. The spectral intensity also varies in a complex way inside the spinon continuum, reflecting scattering between different bands. Throughout this work, we use $b=1.3$, $\lambda=1$, and $\theta=0.45$ to measure the amplitude and phase of the CDW order.

\begin{figure}[H]
\begin{minipage}[h]{1\linewidth}
  \center{\includegraphics[width=1\linewidth]{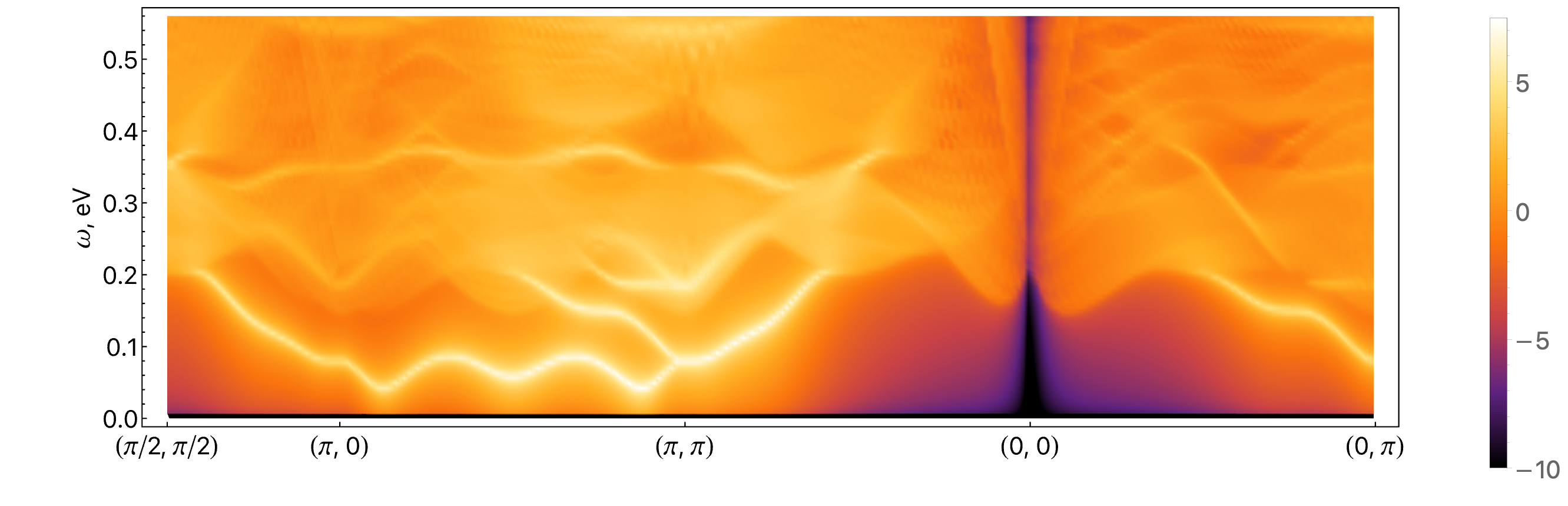}}
  \end{minipage} 
\caption{RPA $\pi$-flux spin structure factor $-\Im \chi^{RPA}(\omega,q)$  on a logarithmic scale in the presence of CDW order, averaged over the $x$ and $y$-directions. RPA corrections to the susceptibility introduce a triplon line-shape of high intensity, having an hourglass-like form near $(\pi,\pi)$.}
\label{fig:susceptibility_piflux_RPA_CDW}
\end{figure}

We examine the effect of the interacting part of the Hamiltonian using the results of the previous section, see Eq.~(\ref{eqn:Ham_int}). Incorporating interactions within the RPA framework leads to a renormalized spin susceptibility given by Eq.~\eqref{eq:susceptibility_RPA}. Fig.~\ref{fig:susceptibility_piflux_RPA_CDW} demonstrates that the including interactions leads to an emergent collective excitation (a triplon branch) below the spinon continuum. We choose Heisenberg couplings $J=0.235,J'=-0.3J,J''=0.6J$,$J'''=0.02J$, $J''''=0.45J$ to achieve the dispersion that reproduces the characteristic hourglass-shaped spectrum. We note that at half-filling, where the system is a Mott insulator, the coupling $J$ is exactly the nearest-neighbor antiferromagnetic Heisenberg exchange coupling, which has been estimated to be about 200 meV. This type of spectrum has been observed in numerous neutron-scattering experiments on cuprate materials\cite{Hayden2004,Vignolle2007,Chan2016}. 

Although the model contains several phenomenological parameters, we note that the hourglass-shaped spectrum can be reproduced using only the $J$, $J'$, and $J''$ couplings. However, in that case, the characteristic features appear at lower energies and are less distinct.
The CDW order affects the Heisenberg couplings as well, see Eq.~(\ref{eqn:Ham_int}), and we chose $g=0.1$ to control the relative amplitude.

\begin{figure}[h!]
\begin{minipage}[h]{1\linewidth}
  \center{\includegraphics[width=1\linewidth]{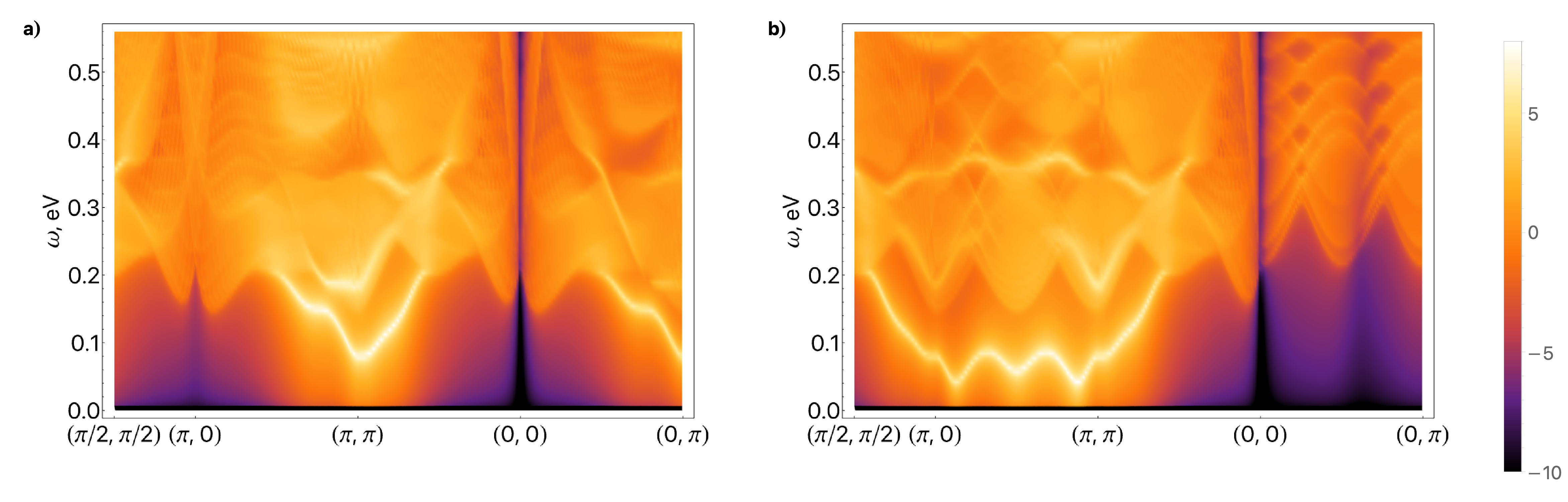}}
  \end{minipage} 
\caption{RPA $\pi$-flux spin structure factor on a logarithmic scale in the presence of CDW order along the $x$-direction (a) and along the $y$ direction (b). The minimum of the triplon branch is centered at $(\pi,\pi)$ when the CDW order is set along the $x$-direction, and it is shifted from  $(\pi,\pi)$ when the order is along the $y$-direction, producing a famous hourglass-like spectrum after averaging over both directions.}
\label{fig:susceptibility_piflux_RPA_CDW_separate}
\end{figure}

Importantly, the hourglass profile arises from averaging the CDW order over both spatial directions. The motivation for this is the following: the correlation length in most cuprate materials is quite small($<~\SI{100 }{\angstrom}$), and in any real sample there are domains with various order orientation \cite{Tranquada1995,Fujita2004}. A similar averaging procedure was done in Ref. \cite{Vojta2004} to study the hourglass shaped spectrum of LBCO.
Without this averaging, two distinct triplon branches appear, as illustrated in Fig.~\ref{fig:susceptibility_piflux_RPA_CDW_separate}. Panels (a) and (b) correspond to the CDW order along the $x$ and $y$ directions correspondingly. In panel (a), the triplon branch exhibits a minimum at $(\pi,\pi)$, whereas in panel (b) the minimum is shifted away from $(\pi,\pi)$. The combination of the two produces the characteristic hourglass-shaped dispersion.

In Appendix~\ref{app:RPA_noCDW} we analyze RPA renormalized spin structure factor  without the CDW order. While the triplon branch is observable, it does not have a typical hourglass shape and the spinon continuum spreads all the way to zero energies, making it harder to relate to the experimental observations.

\begin{figure}[h!]
\begin{minipage}[h]{1\linewidth}
  \center{\includegraphics[width=1\linewidth]{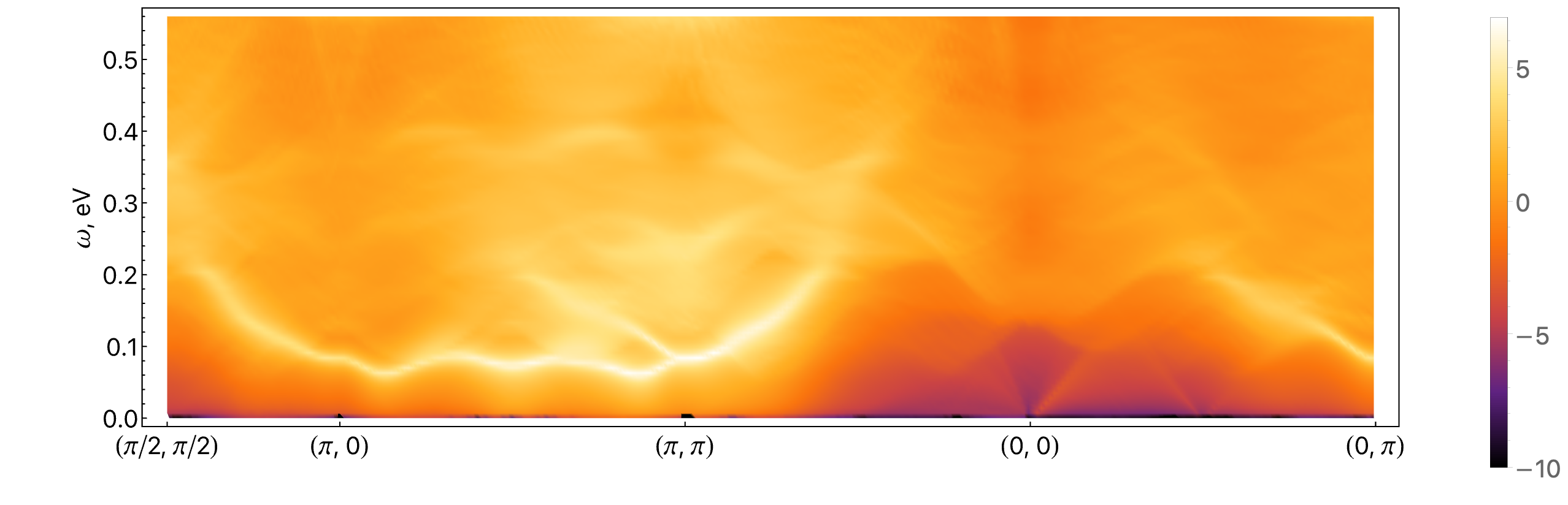}}
  \end{minipage} 
\caption{RPA spin structure factor on a logarithmic scale in the presence of CDW order averaged over the $x$ and $y$-directions, when all three ancilla layers hybridize and contribute to the overall susceptibility. The presence of hole pockets and gapless excitations in the first two layers result in the finite spectral weight inside the CDW gap and broadens the triplon branch. }
\label{fig:susc_piflux_electrons_CDW_RPA_el}
\end{figure}

So far, we have considered only the contribution of the third layer to the spin structure factor, neglecting contributions from the other layers. In principle, however, the full Hamiltonian is given by Eq.~(\ref{eq:hamiltonian_electrons}), and the contribution from the remaining layers is non-negligible. In particular, the first two layers host gapless excitations near the Fermi energy, forming hole pockets, or so-called Fermi arcs. Consequently, their presence is expected to modify the low-energy part of the structure factor, introducing additional spectral weight and damping effects in this regime. Fig.~\ref{fig:susc_piflux_electrons_CDW_RPA_el} shows that at a finite hybridization strength between the third and second layers ($g_e=0.4$) the triplon branch and the spinon continuum become noticeably broadened, while the gapped region in the spectrum is softened.

\begin{figure}[h!]
    \begin{minipage}[h]{1\linewidth}
    \center{\includegraphics[width=0.5\linewidth]{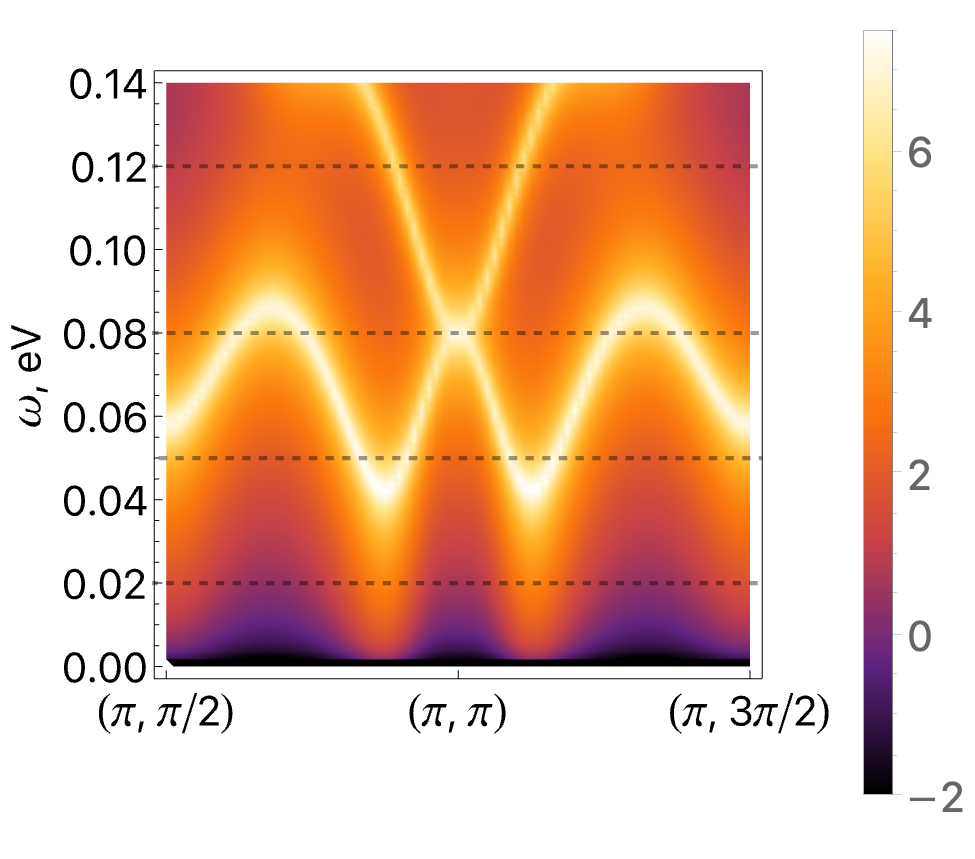}}
    \end{minipage} 
\caption{RPA $\pi$-flux spin structure factor on a logarithmic scale in the presence of CDW order averaged over the $x$ and $y$-directions. The triplon branch exhibits a characteristic hourglass-like shape, centered around $80 \, meV$. Dashed lines represent constant energy cuts along the 2D Brillouin zone in Fig.~\ref{fig:susc_piflux_electrons_CDW_RPA_energy_cuts}.}
\label{fig:plots_piflux_electrons_CDW_RPA_zoom}
\end{figure}

Fig.~\ref{fig:plots_piflux_electrons_CDW_RPA_zoom} shows a zoomed-in view of hourglass shaped dispersion when $q_y$ varies from $\pi/2$ to $3\pi/2$. We observe a clear hourglass shaped triplon excitation, centered around $(\pi,\pi)$. The minimum of the hourglass occurs at the incommensurate momenta slightly displaced from $(\pi,\pi)$. Notably, in contrast to experimental observations, our calculations show that when the momentum is tuned further away from these minima, the excitation branch rises again. This discrepancy may come from limitations of the theoretical model or from various experimental factors that enhance the magnetic response near  $(\pi,\pi)$. 

\begin{figure}[h!]
    \begin{minipage}[h]{1\linewidth}
    \center{\includegraphics[width=1\linewidth]{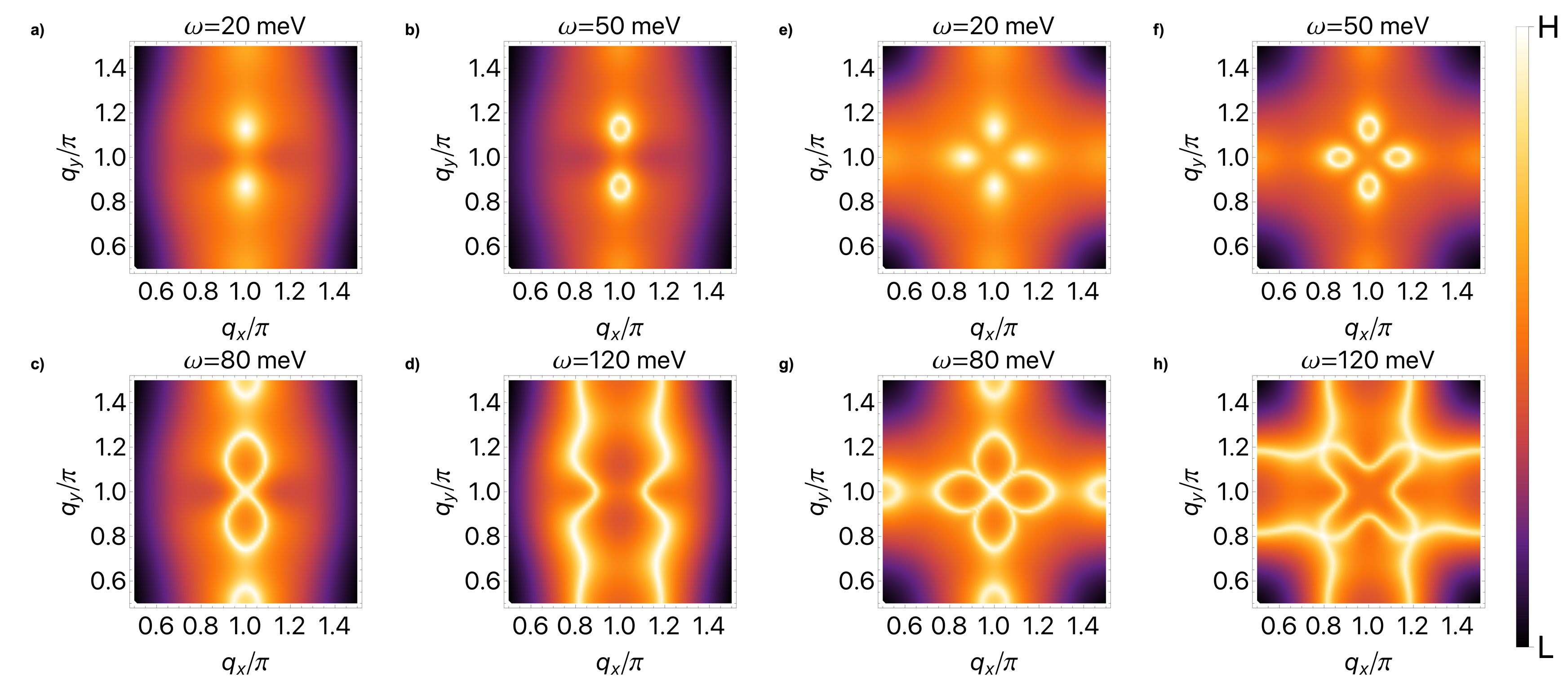}}

    \end{minipage} 
\caption{Energy cuts of RPA $\pi$-flux spin structure factor in the presence of CDW order. Panels (a)-(d) correspond to energies of $20$, $50 $, $80$ and $120 \, meV$ (compare to dashed lines in Fig.~\ref{fig:plots_piflux_electrons_CDW_RPA_zoom}) with the CDW order along the $x$-direction, while panels (e)-(h) correspond to CDW order averaged over the $x$ and $y$-directions, and the resulting plots have $C_4$ symmetry.}
\label{fig:susc_piflux_electrons_CDW_RPA_energy_cuts}
\end{figure}

We also examine constant-energy cuts across the Brillouin zone, with non-averaged (a)-(d) and averaged (e)-(h) CDW order, see Fig.~\ref{fig:susc_piflux_electrons_CDW_RPA_energy_cuts}. In particular, panels (a),(b),(e), and (f) demonstrate that below the neck of the hourglass, the intensity maxima are displaced from $(\pi,\pi)$. Around $\omega\approx 80 \, meV$ a pronounced peak emerges at $(\pi,\pi)$ indicating that this energy corresponds to the center of the hourglass dispersion. At higher energies, the central intensity weakens again and the triplon branch forms a complex pattern.

It is worthwhile to additionally explore the cut along the $(\pi,0)$ direction in the Brillouin zone. The bare spin structure factor shows a strong gap at low energies and broad spinon continuum at higher energies. In contrast, RPA spin structure factor demonstrates an additional peak inside the gap, coming from the triplon branch, while the overall intensity of the spinon continuum is reduced. 

 \begin{figure}[h!]
    \begin{minipage}[h]{0.45\linewidth}
    \center{\includegraphics[width=1\linewidth]{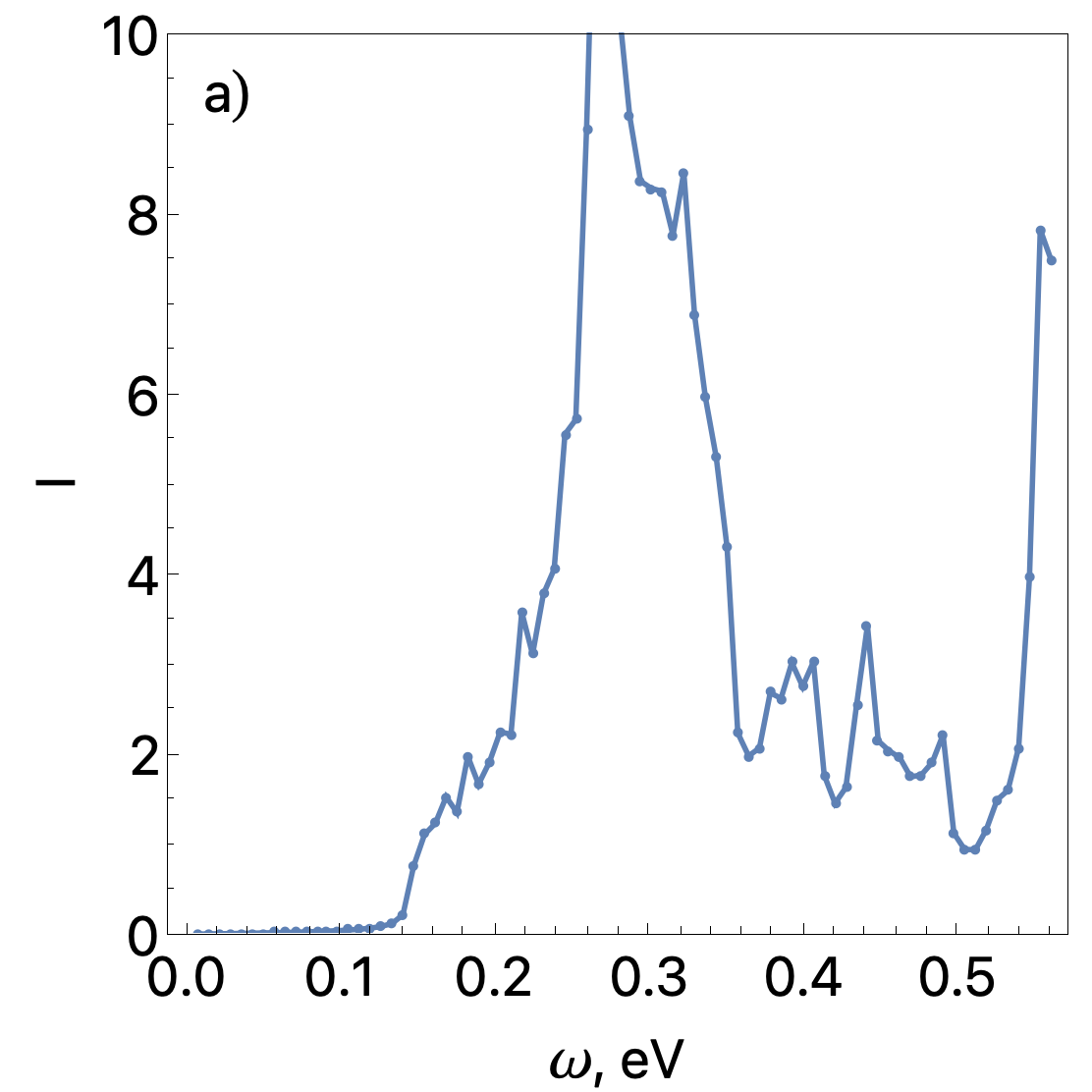}}
    \end{minipage} 
            \hfill
     \begin{minipage}[h]{0.45\linewidth}
    \center{\includegraphics[width=1\linewidth]{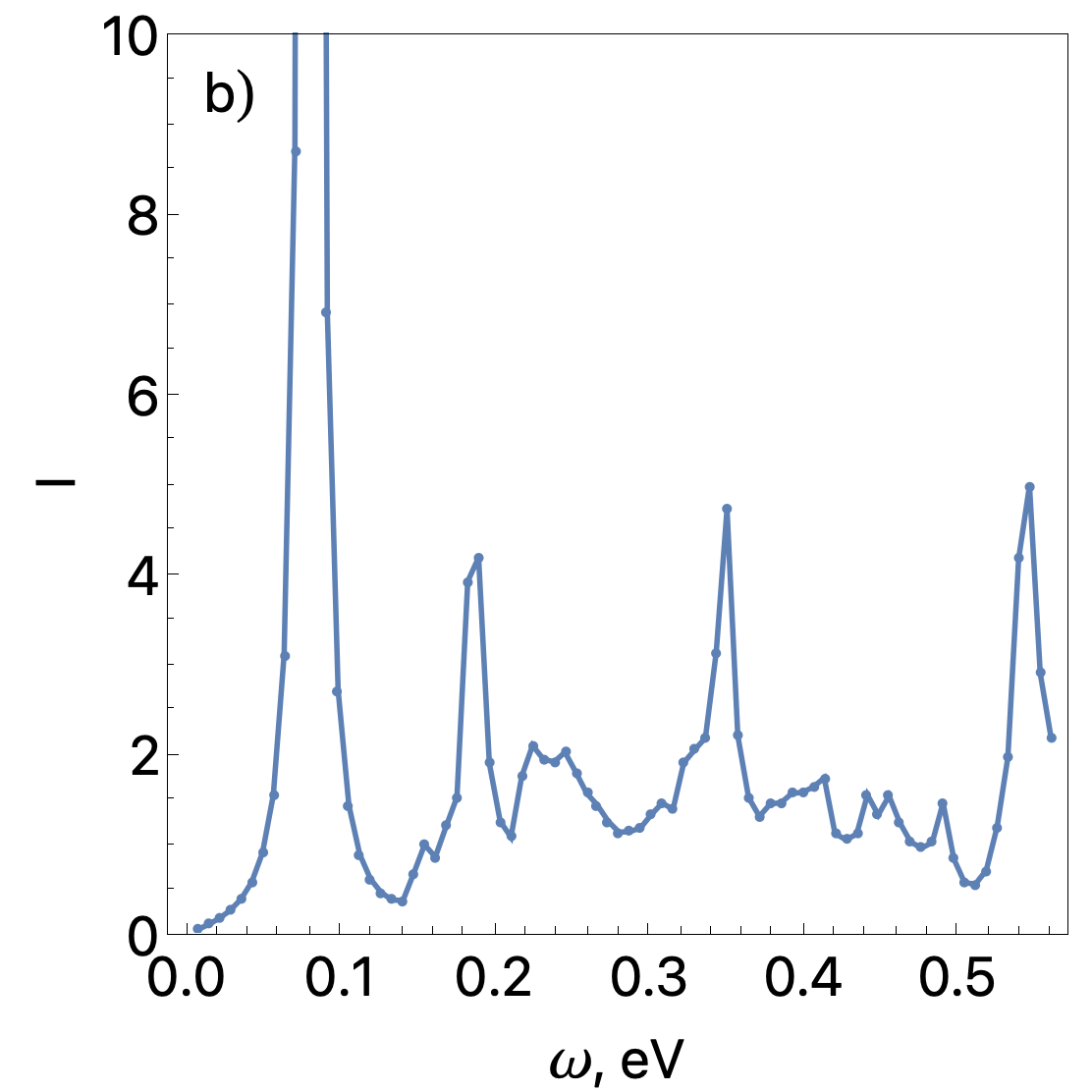}}    
    \end{minipage} 
\caption{1D bare (a) and RPA (b) $\pi$-flux spin structure factor at fixed momentum $(q_x,q_y)=(\pi,0)$ in the presence of CDW order averaged over the $x$ and $y$ directions. Panel (a) shows a spinon continuum starting at $140 \, meV$, while panel (b) shows an additional high-intensity triplon branch around $80 \,meV$ and a redistribution of the spectral weight inside the spinon continuum.  }
\label{fig:plots_piflux_electrons_CDW_RPA_2}
\end{figure}

Motivated by neutron-scattering experiments in a magnetic field~\cite{Coldea1997}, we calculate the spin susceptibility in our model in the presence of a finite magnetic field in Appendix~\ref{app:magnetic}. We find that the hourglass-like triplon branch splits into two branches, which are shifted in opposite directions by an energy of $2\Delta E_z$  due to Zeeman coupling. Although experiments require strong magnetic fields to resolve this splitting, such conditions may become possible in the future.

Finally, we comment on an alternative theoretical approach capable of reproducing the hourglass-shaped magnetic response, discussed in Appendix~\ref{app:altern}. In particular, we employ the phenomenological theory of coupled spin and charge fluctuations introduced in Ref.~\cite{Vojta2006}, to obtain an hourglass-shaped triplon branch outside the spinon continuum. Although this approach is valuable for its simplicity, the RPA-based susceptibility calculated in the main text offers a more direct and microscopic route to reproduce the experimental observations.
\section{Conclusions}
\label{sec:conclusions}

We have provided a detailed view of the spin structure factor in the ground state of a doped antiferromagnet with period 4 charge density wave order. It is assumed that the intermediate scale spin physics is that of the a particular spin liquid, as is required for consistency with the structure of the intermediate temperature pseudogap. The spinons eventually confine at low energies, by the condensation of a Higgs field, $B$, which produces the CDW order and simultaneously quenches the emergent gauge field. 

Near the charge-ordering wavevector, a RPA analysis of the interaction between the spinons produces the `hourglass' spectrum observed by neutron scattering \cite{Hayden2004,Vignolle2007,Chan2016}. Above the hourglass we obtain additional scattering from the spinon continuum which we hope will be studied in subsequent experiments. 

The spinons are also visible in at other momenta, both in the form of RPA bound states and two-particle continua. This spectrum is a plausible explanation \cite{SSZaanen} for the high energy scattering observed by RIXS experiments \cite{Keimer11,Hayden19,Keimer25}, and is present at similar energies. 
But the detailed form of our high energy spectrum appeares different from current observations {\it e.g.\/} the high intensity scattering disperses downwards from $(\pi,0)$ to $(0, 0)$ in the RIXS observations \cite{Hayden19,Keimer11}, while that in Fig.~\ref{fig:susceptibility_piflux_RPA_CDW} disperses upwards. We believe this is a consequence of the rather simple condensed form of the Higgs field, $B$, in a period 4 CDW state in our computation. It will be of interest to extend our study to the case of fluctuating CDW states using $B$ and SU(2) gauge fields, along the lines of Ref.~\cite{Sayantan25}. 

We close by summarizing some of the recent arguments for the FL* state with fermionic spinons in the under-hole-doped cuprates:
\begin{itemize}
 \item The FL* theory accounts for the dispersion of the gapped electronic excitations near momentum $(\pi, 0)$ (the `antinode'), with the spinons contributing to the broad linewidth \cite{Mascot2022}.
   \item The existence of the FL* state allows the possibility of an underlying quantum phase transition to a Fermi liquid (FL) without any symmetry-breaking order parameter \cite{FLS1,Zhang2020}. Both FL* and FL have instabilities to the same $d$-wave superconductor, and so there is no quantum criticality within the superconducting state. But a strange metal state appears above the superconducting $T_c$ from FL*-FL quantum criticality in the presence of random interactions \cite{Patel2}: such a state matches transport \cite{Li:2024kxr}, thermopower \cite{LPS24}, and Hall angle \cite{Davide25} observations. 
    \item The nodal quasiparticles of the low temperature $d$-wave superconducting state have strongly anisotropic velocities, with $v_F \gg v_\Delta$, where $v_F$ is the velocity along the zone diagonal, while $v_\Delta$ is the velocity along the orthogonal direction. This is difficult to understand from a $d$-wave superconductor descending from a spin liquid with massless Dirac fermion spinons \cite{Zhang88,KotliarLiu88,IvanovSenthil,LeeWenRMP}, as the spinons have nearly isotropic velocities. However, anisotropic nodal quasiparticles are obtained naturally by confining the FL* state with such a spin liquid background \cite{Chatterjee16,Christos2024}: the spinons annihilate the Bogoliubov quasiparticles on the `backsides' of the pockets, and observed Bogoliubov quasiparticles of the $d$-wave superconductor are those on the `front side' of the pockets. 
     \item The hole-doped quasiparticles display quantum oscillations at low temperatures and high magnetic fields. These oscillations are associated with {\it electron\/} pockets, as indicated by the negative Hall co-efficient in this regime. These oscillations are compatible with a model of electron pockets induced by bi-directional charge density wave order \cite{HS11} only after accounting for the influence of spinons \cite{BCS24}.
    \item The observed structure of the vortex core in the underdoped cuprates is clearly distinct from that of the BCS theory of a $d$-wave superconductor emerging from a Fermi liquid \cite{WM95}. The key differences can be described by a theory of the BCS $d$-wave superconductor emerging from a confinement transition of the FL* state \cite{ZhangSS24,Sayantan25}.
    \item The hole pockets survive in a $T=0$ FL* state with a non-zero quasiparticle residue around the hole pocket, even in the presence of the coupling to spinons \cite{Mascot2022}. Recent work \cite{Sayantan25} examined thermal fluctuations of a SU(2) gauge theory describing the spinons, at temperatures above a superconducting or charge-ordered ground state (the spinons are confined in these low temperature states). They showed that such thermal fluctuations do indeed convert the photoemission spectral weight to that of the observed `Fermi arcs' in the intermediate temperature pseudogap regime.
    \item Finally, we note the evidence from ADMR and the Yamaji effect \cite{fang_admr_2022,chan_yamaji_2024,Zhao_Yamaji_25}, which was discussed in Section~\ref{sec:intro}. 
\end{itemize}

\subsection*{Acknowledgements}

This research was supported by NSF Grant DMR-2245246 and by the Simons Collaboration on Ultra-Quantum Matter which is a grant from the Simons Foundation (651440, S. S.). 
P.M.B. acknowledges support by the German National Academy of Sciences Leopoldina through Grant No.~LPDS 2023-06 and the Gordon and Betty Moore Foundation’s EPiQS Initiative Grant GBMF8683.
The Flatiron Institute is a division of the Simons Foundation.

\appendix

\section{Theoretical details}
\label{app:CDW}
In the basis $F_{2,\bm{q}}=(f_{2,\bm{q}},f_{2,\bm{q}+K_x},f_{2,\bm{q}+2 K_x},f_{2,\bm{q}+3 K_x})$, the Hamiltonian in Eq. \eqref{eq:H_pi_flux_CDW} becomes a $4\times 4$ matrix:
\begin{equation}
H_{f_2,f_2}= -t^{f_2} F_{2,\bm{q}}^\dag
\left(
    \begin{array}{cccc}
       \varepsilon_0(\bm{q})&\varepsilon_1(\bm{q})&\varepsilon_2(\bm{q})&\varepsilon^*_1(\bm{q}+3K_x)\\
       \varepsilon_1(\bm{q})^*&   \varepsilon_0(\bm{q}+K_x)&\varepsilon_1(\bm{q}+K_x)&\varepsilon_2(\bm{q}+K_x)\\
       \varepsilon_2(\bm{q})^*&\varepsilon_1(\bm{q}+K_x)^*&    \varepsilon_0(\bm{q}+2K_x)&\varepsilon_1(\bm{q}+2K_x)\\
       \varepsilon_1(\bm{q}+3K_x)&\varepsilon_2(\bm{q}+K_x)^*&\varepsilon_1(\bm{q}+2K_x)^*&   \varepsilon_0(\bm{q}+3K_x)\\
    \end{array}
    \right)    F_{2,\bm{q}},
\label{eq:hamiltonian_CDW}
\end{equation}

with the matrix elements:

\begin{equation}
\begin{split}    
  & \varepsilon_0(\bm{q})=2 \sin q_x(1+2 \lambda b^2 \cos 2 \theta  \cos (K_x/2))+4 \lambda b^2   \sin q_y,\\
  &\varepsilon_1(\bm{q})=i \lambda b^2  \cos 2 \theta  (e^{-i q_x- i \phi-i K_x/2}- e^{iq_x- i \phi+i K_x/2})+2 \lambda b^2 \sin 2 \theta \sin q_y e^{i \phi}+2 \lambda b^2   \sin q_y e^{-i \phi},\\
  & \varepsilon_2(\bm{q})=2\sin q_y(1+2\lambda b^2 \sin 2 \theta) .\\
\end{split}
\end{equation}

To go to a different basis $\tilde{F}_{\bm{q}}=(f_{2,\bm{q},\delta_1},f_{2,\bm{q},\delta_2},f_{2,\bm{q},\delta_3},f_{2,\bm{q},\delta_4})$ with $\delta_i$ being a sublattice site, one can use the unitary transformation:
\begin{equation}
 H_{\delta_i\delta_j}(\bm{q})=U_{\delta_i n}H_{nm}(\bm{q}) U^\dag_{m \delta_j}, \quad \quad U_{a b}=(1/2)e^{i K_x a b}
\end{equation}

The Heisenberg coupling is represented by a matrix $\tilde{J}$. In the basis $F_{\bm{q}}=(f_{\bm{q}},f_{\bm{q}+K_x},f_{\bm{q}+2 K_x},f_{\bm{q}+3 K_x})$, it reads
\begin{equation}
\tilde{J}=F_{n}^*(\bm{q})J_{nm}F_m=J F_{\bm{q}}^*
\left(
    \begin{array}{cccc}
       J_0(\bm{q})&J_1(\bm{q})&J_2(\bm{q})&J_1(\bm{q}+3K_x)^*\\
       J_1(\bm{q})^*&   J_0(\bm{q}+K_x)&J_1(\bm{q}+K_x)&J_2(\bm{q}+K_x)\\
       J_2(\bm{q})^*&J_1(\bm{q}+K_x)^*&    J_0(\bm{q}+2K_x)&J_1(\bm{q}+2K_x)\\
       J_1(\bm{q}+3K_x)&J_2(\bm{q}+K_x)^*&J_1(\bm{q}+2K_x)^*&   J_0(\bm{q}+3K_x)\\
    \end{array}
    \right)    F_{\bm{q}},
    \label{eq:J_coupling}
\end{equation}

\begin{equation}
\begin{split}    
  & J_0(\bm{q})=2\cos q_x \left( 1+2 g b^2 \cos 2 \theta \cos(K_x/2)\right)+2\cos q_y(1+2g b^2 \sin 2 \theta) ,\\
  &J_1(\bm{q})=g b^2 \cos 2 \theta  (e^{-i q_x- i \phi-i K_x/2} +e^{iq_x-i \phi+i K_x/2 })+2  gb^2 \sin 2 \theta\cos q_y e^{-i \phi}+2 g b^2 \cos q_y e^{i \phi},\\
  & J_2(\bm{q})=4g b^2  \cos q_y,\\
\end{split}
\end{equation}

We can also add the second,third, and fourth nearest neighbor coupling to reproduce the hourglass features: $\tilde{J}_0(\bm{q})=J_0(\bm{q})+4J' \cos q_x \cos q_y+2J'' (\cos2q_x+\cos 2 q_y)+4J'''(\cos2q_x \cos q_y+\cos q_x\cos 2 q_y)$.

To produce Fig.~\ref{fig:susc_piflux_electrons_CDW_RPA_el}, we account for the coupling between the $\pi$-flux state and other layers of Ancilla model. The full three-layer Hamiltonian is defined in the basis ($C_{\bm{q}},F_{1,\bm{q}},F_{2,\bm{q}}$), where each operator corresponds to a certain level. In the presence of CDW order, all operators are defined in the reduced Brillouin zone $C_{\bm{q}}=(c_{\bm{q}},c_{\bm{q}+K_x},c_{\bm{q}+2K_x},c_{\bm{q}+3K_x})$
and similarly for $F_{1,\bm{q}},F_{2,\bm{q}}$. 
The full Hamiltonian consists of three blocks: 
$H_{c,c}(\bm{q})=diag(\epsilon_{c}(\bm{q}),\epsilon_{c}(\bm{q}+K_x),\epsilon_{c}(\bm{q}+2K_x),\epsilon_{c}(\bm{q}+3K_x))$ and $H_{f_1,f_1}$ has the same structure with $\epsilon_f$ instead of $\epsilon_c$. The coupling between the first and second layers is described via a hybridization field $H_{c, f_1}=\Phi I_{4\times 4}$.
The block $H_{f_2,f_2}$ corresponds to the Hamiltonian of the $\pi$-flux state introduced earlier. The $B$ field is defined according to Eq.~\eqref{eq:B_field}. For convenience, we redefine the operator $\tilde{f_2}^\dag(\bm{k})=e^{i \phi_2}f_2^\dag(k_x-K_x/2+\pi/2,k_y+\pi/2) $ to absorb the prefactor appearing in $B$. The matrix form of $B$ reads:
\begin{equation}
i g_e B= i b g_e
\left(
    \begin{array}{cccc}
S_1&S_2e^{i \phi}& S_2 & S_1 e^{i \phi}\\
S_1 e^{ i \phi} & S_1&S_2e^{i \phi}&S_2\\
S_2 &S_1 e^{ i \phi}&S_1&S_2e^{i \phi}\\
S_2e^{i \phi}&S_2&S_1 e^{ i \phi}&S_1
    \end{array}
    \right)  ,
\end{equation}

The Hamiltonian block $H_{f_2,f_2}$ should also be modified
\begin{equation}
H_{f_2,f_2}(\bm{q})= t^{f_2}(F_2)_{\bm{q}}^*
\left(
    \begin{array}{cccc}
       \varepsilon_0(\bm{q})&0&\varepsilon_2(\bm{q})&0\\
       0&   \varepsilon_0(\bm{q}+K_x)&0&\varepsilon_2(\bm{q}+K_x)\\
       \varepsilon_2(\bm{q})^*&0&    \varepsilon_0(\bm{q}+2K_x)&0\\
       0&\varepsilon_2(\bm{q}+K_x)^*&0&   \varepsilon_0(\bm{q}+3K_x)\\
    \end{array}
    \right)    (F_2)_{\bm{q}},
\end{equation}

\begin{equation}
 \varepsilon_0(\bm{q})=2 \sin(q_x+\pi/4),\quad\varepsilon_2(\bm{q})=2\sin(q_y+\pi/2).
\end{equation}

After diagonalizing the Hamiltonian, we use Eq.~(\ref{susc:electrons}) to compute the susceptibility. It can be seen from Eq.~\eqref{susc:electrons} that the susceptibility is invariant with respect to going from $f_2(\bm{k}) $ to the $\tilde{f}_2(\bm{k})$ basis. 

\section{RPA $\pi$-flux}
\label{app:RPA_noCDW}
\begin{figure}[h!]
\begin{minipage}[h]{1\linewidth}
  \center{\includegraphics[width=1\linewidth]{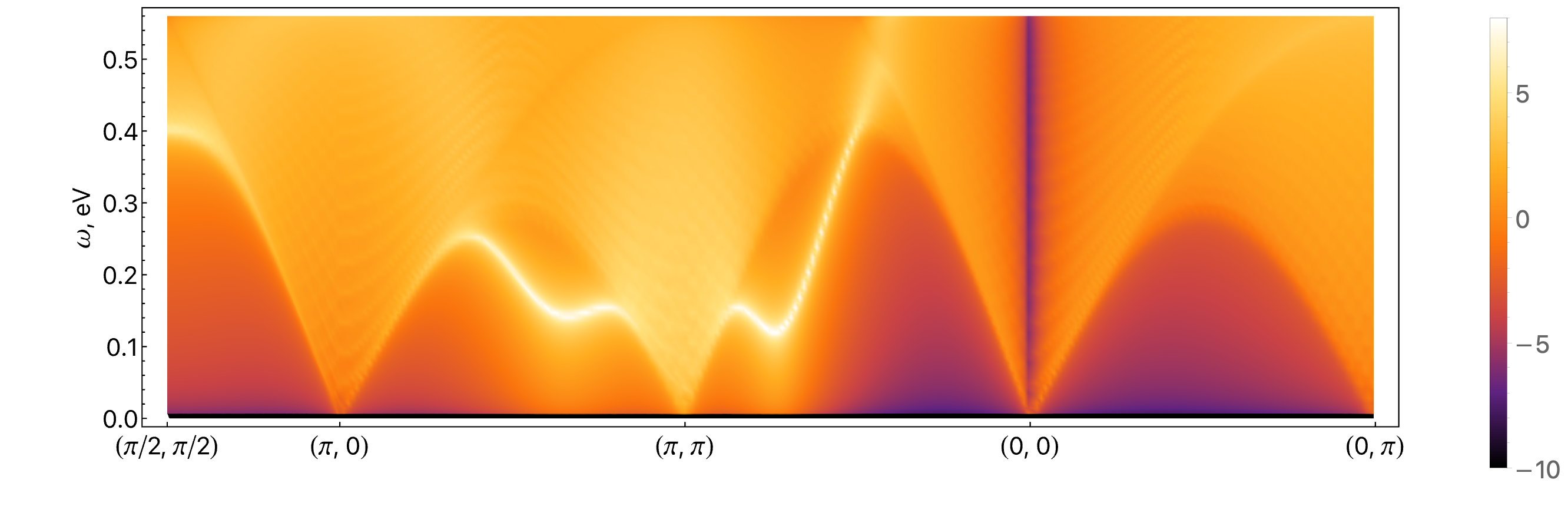}}
  \end{minipage} 
\caption{RPA $\pi$-flux spin structure factor, shown on a logarithmic scale, without CDW order. The spinon continuum retains a Dirac-like shape near $(\pi,\pi)$, while a triplon branch at high energies merges with the spinon continuum near $(\pi,\pi)$.}
\label{fig:piflux_RPA_noCDW}
\end{figure}

In this Appendix we compute RPA susceptibility of the $\pi$-flux state without the CDW order, see Fig.~\ref{fig:piflux_RPA_noCDW}. The spectrum remains gapless and Dirac-like spinon continuum near $(\pi,\pi)$ is still visible. An additional triplon branch emerges at higher energies and merges with the spinon continuum near $(\pi,\pi)$. Therefore, including the CDW is crucial in obtaining the hourglass-like spin-structure factor.

\section{Adding magnetic field}
\label{app:magnetic}
In this Appendix we analyze how the spin structure factor changes in the presence of a magnetic field, applied perpendicularly to the sample. The energy bands experience a Zeeman splitting, since the Hamiltonian becomes $H_B=H_0+\mu B S^z$. Spin susceptibilities are defined in the standard way:
\begin{equation}
    \chi_{zz}(\omega,q)=\frac{1}{2} \left(\chi_{\uparrow \uparrow}(\omega,q)+\chi_{\downarrow \downarrow}(\omega,q) \right) \quad \quad   \chi_{+ -}(\omega,q)=\chi_{\downarrow \uparrow}(\omega,q) .
\end{equation}

The $\chi_{\uparrow \uparrow}$ component of susceptibility could be written by generalizing Eq.~(\ref{eq:susc:CDW}):

\begin{equation}
    \chi^0_{\delta_i,\delta_j,(\uparrow \uparrow)}(\omega,\bm{q})=\sum_{\alpha,\beta=1,..,4}\sum_{\bm{k}} \frac{n_F(E_\alpha(\bm{k})+\Delta E_z)-n_F(E_\beta(\bm{k}+\bm{q})+\Delta E_z)}{\omega+i \delta+E_\alpha(\bm{k})-E_\beta(\bm{k}+\bm{q})}  F^\alpha_{\delta_i}(\bm{k}) F^{\alpha}_{\delta_j}(\bm{k})^*  F^\beta_{\delta_j}(\bm{k}+\bm{q})F^\beta_{\delta_i}(\bm{k}+\bm{q})^* ,
    \label{eq:susc:CDW_zeeman_upup}
\end{equation}
, where $\Delta E_z$ is the Zeeman splitting. Since the scattering occurs between the same spin species, shifted by the same energy, $\chi_{\uparrow \uparrow}$ component of susceptibility is not affected by the magnetic field. The $\chi_{\uparrow \downarrow}$ component is defined in the following way:

\begin{equation}
    \chi^0_{\delta_i,\delta_j,(\uparrow \downarrow)}(\omega,\bm{q})=\sum_{\alpha,\beta=1,..,4}\sum_{\bm{k}} \frac{n_F(E_\alpha(\bm{k})+\Delta E_z)-n_F(E_\beta(\bm{k}+\bm{q})-\Delta E_z)}{\omega+i \delta+E_\alpha(\bm{k})-E_\beta(\bm{k}+\bm{q})+2\Delta E_z}  F^\alpha_{\delta_i}(\bm{k}) F^{\alpha}_{\delta_j}(\bm{k})^*  F^\beta_{\delta_j}(\bm{k}+\bm{q})F^\beta_{\delta_i}(\bm{k}+\bm{q})^* .
    \label{eq:susc:CDW_zeeman_updown}
\end{equation}

\begin{figure}[h!]
\begin{minipage}[h]{1\linewidth}
  \center{\includegraphics[width=0.5\linewidth]{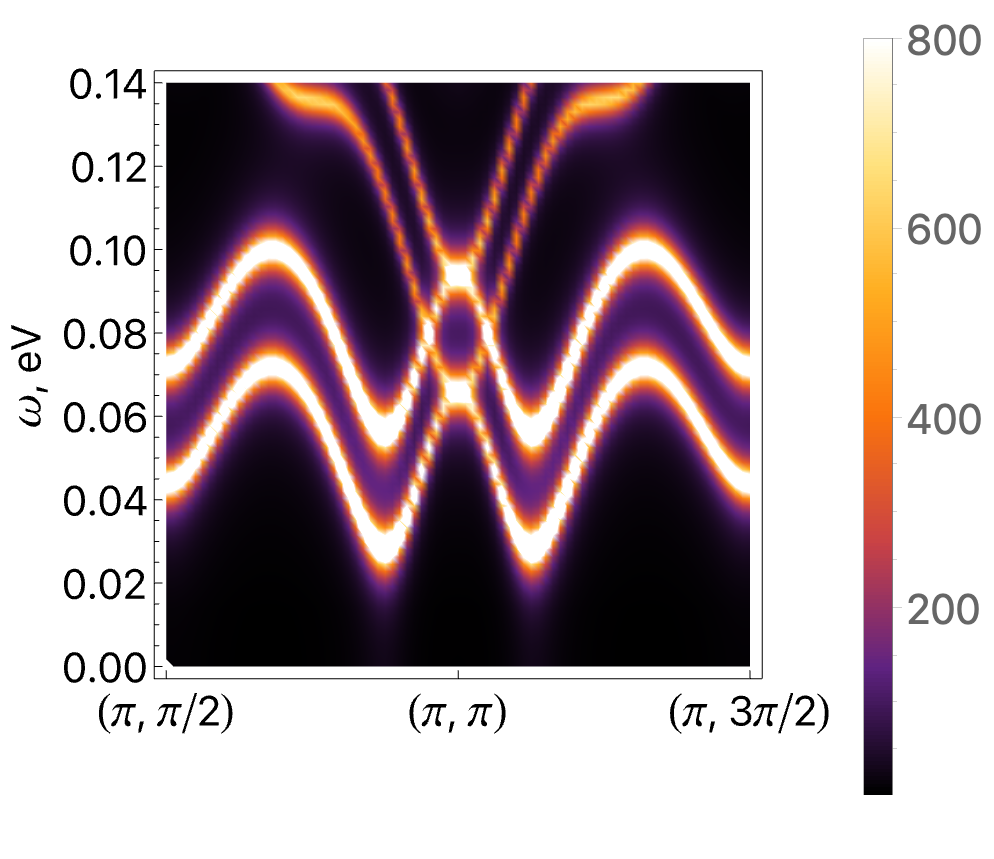}}
  \end{minipage} 
\caption{RPA $\pi$-flux spin structure factor  $-\Im \left(\chi_{+-}(\omega,q)+\chi_{-+}(\omega,q) \right)$ in the presence of a magnetic field $B=60 \,T$ with a Zeeman splitting $\Delta E_z\approx0.007 \, eV$. Two triplon branches with $S_z=\pm1$ are shifted by $\pm2\Delta E_z$ in energy. }
\label{fig:plots_piflux_magnetic}
\end{figure}

The $\chi_{\uparrow \downarrow}$ component of susceptibility accounts for scattering between opposite spin orientations, therefore, the resulting triplon branch is shifted by $2\Delta E_z$. Similarly, the triplon branch coming from $\chi_{-+}$ component would be shifted by $2\Delta E_z$ in the opposite direction, as shown in Fig.~\ref{fig:plots_piflux_magnetic}. In practice the splitting is hard to observe since it requires both good energy resolution and strong magnetic fields. However, our prediction could guide future experiments in confirming the triplon spin character of the hourglass-shaped branch.

\section{Alternative approach}
\label{app:altern}
In the main part of the paper we obtained the hourglass-like triplon branch as a collective mode of the RPA susceptibility. In this Appendix,  we demonstrate that similar features exist in a model of fluctuating stripe charge order, proposed in \cite{Vojta2006}. We use a phenomenological Lagrangian to describe the coupling between spin and charge fluctuations, which takes the following form:

\begin{equation}
    \mathcal{L}=\rho_i \varphi_i^2+\lambda_2 (\varphi_i \varphi_{i+x}+\varphi_i \varphi_{i+y})+Q_{b,ij}\varphi_i \varphi_j,
\end{equation}
where $\varphi$-field describes the spin fluctuations and $Q$-field measures the fluctuations of charge density. Similar to the main text, we choose CDW order with a period of four unit cells:
\begin{equation}
    \rho_i=\lambda_1+2 \lambda_3 \cos \left(\frac{\pi}{2} x +\phi\right), \quad \quad Q_{b,i-x,i+x}=Q_{b,i-y,i+y}=2 \lambda_4\cos \left( \frac{\pi}{2}x+2\phi\right).
\end{equation}
The phase $\phi=\pi/4$ ensures that CDW order is bond centered. We define a reduced unit cell operator $\varPhi_q=(\varphi_q,\varphi_{q+\pi/2},\varphi_{q+\pi},\varphi_{q+3\pi/2})$, with the period $K_x=(\pi/2,0)$. In this basis, the Hamiltonian is

\begin{equation}
H(q)= t^{f_2}\varPhi_q^*
\left(
    \begin{array}{cccc}
       \varepsilon_0(q)&\varepsilon_1(q)&0&\varepsilon^*_1(q+3K_x)\\
       \varepsilon_1(q)^*&   \varepsilon_0(q+K_x)&\varepsilon_1(q+K_x)&0\\
       0&\varepsilon_1(q+K_x)^*&    \varepsilon_0(q+2K_x)&\varepsilon_1(q+2K_x)\\
       \varepsilon_1(q+3K_x)&0&\varepsilon_1(q+2K_x)^*&   \varepsilon_0(q+3K_x)\\
    \end{array}
    \right)    \varPhi_q,
\end{equation}
where the components of the Hamiltonian are

\begin{equation}
\begin{split}    
  & \varepsilon_0(q)=\lambda_1+\lambda_2 (\cos  q_x+\cos  q_y)\\
  &\varepsilon_1(q)=\lambda_3 e^{-i \phi}+\lambda_4  (e^{-i\left(2q_x+\frac{\pi}{2}+\phi \right)}+e^{-i( 2q_y+\phi)}).
\end{split}
\end{equation}

\begin{figure}[h!]
\begin{minipage}[h]{1\linewidth}
  \center{\includegraphics[width=1\linewidth]{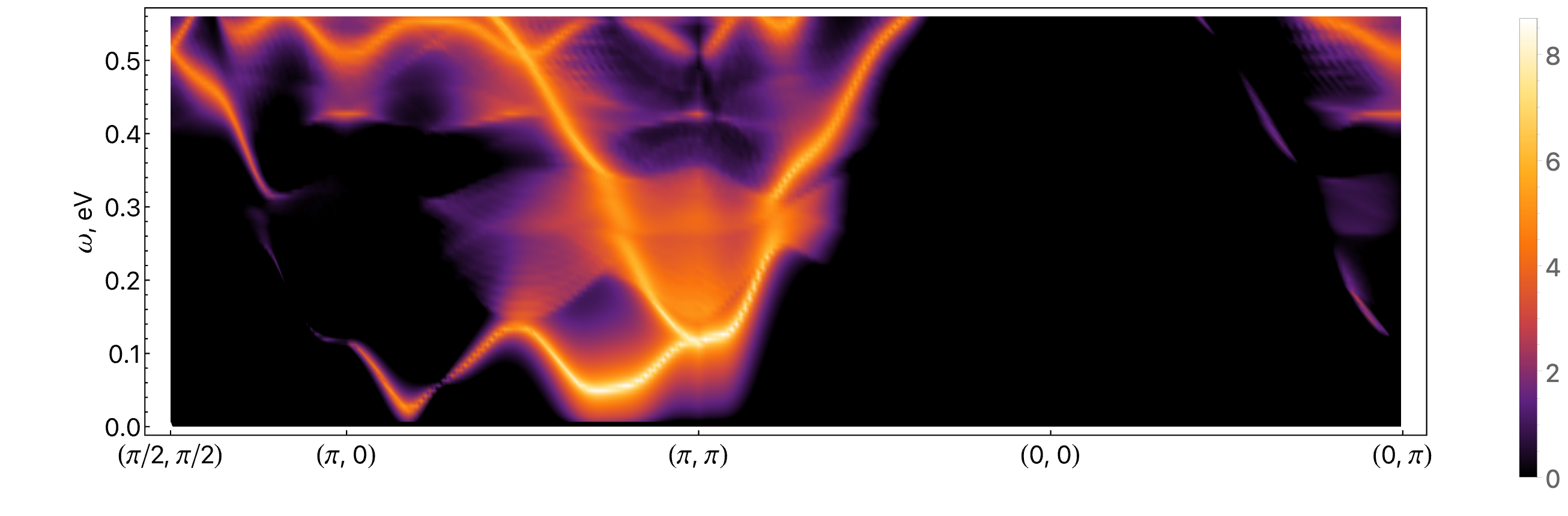}}
  \end{minipage} 
\caption{Imaginary part of the retarded Green's function of spin fluctuations, shown on a logarithmic scale. Parameters $\lambda_1=4.15$, $\lambda_2=1.54$, $\lambda_3=0.07$, $\lambda_4=0.66$. The spectral function demonstrates both a spinon continuum and an hourglass-like triplon branch originating from the spin fluctuations.}
\label{fig:hourglass}
\end{figure}

The eigenenergies satisfy $\omega^2 \varPhi=H(q) \varPhi$ and the Green's function of spin fluctuations is:
\begin{equation}
    G^R(\omega,q)_0=((\omega+i \delta)^2-H(q))^{-1}.
\end{equation}
We also assume that in the real material there is CDW order in both $x$ and $y$ directions, so for the experimental purposes we compute averaged Green's function $\bar{G}(\omega,q_x,q_y)=(G(\omega,q_x,q_y)+G(\omega,q_y,q_x))/2$. The spectral function is defined as $A_{11}(\omega,q)=\Im \bar{G}_{11}(\omega,q_x,q_y)$. 
The $\pi$-flux spinons modify the Green's function in the RPA fashion:
\begin{equation}
    G^R(\omega,q)=((\omega+i \delta)^2-H(q)-g^2\chi(\omega,q))^{-1},
\end{equation}
where $\chi(\omega,q)$ is the $\pi$-flux susceptibility in the presence of CDW order, introduced in Eq.~(\ref{eq:susc:CDW}). 
Fig.~\ref{fig:hourglass} shows the resulting spectral function. We see the hourglass-like dispersion near $(\pi,\pi)$ coming from the effective Hamiltonian and the spinon continuum at higher energies arising from the RPA corrections.

\bibliographystyle{JHEP_new}
\bibliography{main}
\end{document}